\documentclass[aps,prl,twocolumn,superscriptaddress,longbibliography,nobibnotes]{revtex4-2}

\usepackage{diagbox}
\usepackage{amsmath}
\usepackage{amsfonts}
\usepackage{physics}
\usepackage[space]{grffile}
\usepackage{graphicx} 
\usepackage{amssymb}
\usepackage{upgreek}
\usepackage{textgreek}
\usepackage{lineno}
\usepackage{natbib}
\usepackage{braket}
\usepackage{float}
\usepackage{color}
\usepackage{blindtext}
\usepackage{comment}
\usepackage{bm}
\usepackage{lipsum, babel}
\setcitestyle{super}

\usepackage{xr-hyper}
\usepackage{hyperref}
\usepackage{xcolor}
\hypersetup{colorlinks,breaklinks,
            urlcolor=[rgb]{0,0,0.64},
            linkcolor=[rgb]{0,0,0.64},
            citecolor=[rgb]{0,0,0.64},
            filecolor=[rgb]{0,0,0.64}}

\usepackage[T1]{fontenc}

\makeatletter
\renewcommand{\thefigure}{\textbf{\@arabic\c@figure}}
\makeatother

\newcommand{\MIT}{Massachusetts Institute of Technology, Department of Physics, Cambridge, Massachusetts 02139, USA.}
\newcommand{\StanfordAP}{Department of Applied Physics, Stanford University, Stanford, California 94305, USA.}
\newcommand{\StanfordPAP}{Departments of Physics and of Applied Physics, Stanford University, Stanford, California 94305, USA.}
\newcommand{\SIMES}{Stanford Institute for Materials and Energy Sciences, SLAC National Accelerator Laboratory, Menlo Park, California 94025, USA.}

\newcommand{\TDL}{Tsung-Dao Lee Institute, School of Physics and Astronomy, and Zhangjiang Institute for Advanced Study, Shanghai Jiao Tong University, Shanghai 200240, China.}
\newcommand{\TDLI}{Tsung-Dao Lee Institute and School of Physics and Astronomy, Shanghai Jiao Tong University, Shanghai 200240, China.}
\newcommand{\ICQM}{International Center for Quantum Materials, School of Physics, Peking University, Beijing 100871, China.}
\newcommand{\SLAC}{SLAC National Accelerator Laboratory, Menlo Park, CA 94025, USA.}
\newcommand{\IOP}{Beijing National Laboratory for Condensed Matter Physics and Institute of Physics, Chinese Academy of Sciences, Beijing 100190, China.}
\newcommand{\BAQIS}{Beijing Academy of Quantum Information Sciences, Beijing 100913, China.}
\newcommand{\Cornell}{CHESS, Cornell University, Ithaca, New York 14853, USA.}

\begin{document}

\title{Large moir\'{e} superstructure of stacked incommensurate charge density waves}

\author{B.~Q.~Lv}
\thanks{These authors contributed equally: B.Q.L., Y.S., and A.Z.}
\affiliation{\TDL}
\affiliation{\MIT}
\author{Yifan~Su}
\thanks{These authors contributed equally: B.Q.L., Y.S., and A.Z.}
\affiliation{\MIT}
\author{Alfred~Zong}
\thanks{These authors contributed equally: B.Q.L., Y.S., and A.Z.}
\affiliation{\MIT}
\affiliation{\StanfordPAP}
\affiliation{\SIMES}
\author{Qiaomei~Liu}
\affiliation{\ICQM}
\author{Dong~Wu}
\affiliation{\BAQIS}
\author{Noah~F.~Q.~Yuan}
\affiliation{\TDLI}
\author{Zhengwei~Nie}
\affiliation{\IOP}
\author{Jiarui~Li}
\affiliation{\SLAC}
\affiliation{\StanfordAP}
\author{Suchismita Sarker}
\affiliation{\Cornell}
\author{Sheng~Meng}
\affiliation{\IOP}
\author{Jacob P. C. Ruff}
\affiliation{\Cornell}
\author{N.~L.~Wang}
\affiliation{\BAQIS}
\affiliation{\ICQM}
\author{Nuh~Gedik}
\email[Correspondence to: ]{gedik@mit.edu}
\affiliation{\MIT}

\date{\today}

\begin{abstract}
Recent advances in van der Waals heterostructures have opened the new frontier of moir\'{e} physics, whereby tuning the interlayer twist angle or adjusting lattice parameter mismatch have led to a plethora of exotic phenomena such as unconventional superconductivity and fractional quantum spin Hall effect. We extend the concept of moir\'{e} engineering to materials that host incommensurate orders, where we discovered a long-period, thermally-hysteretic moir\'{e} superlattice in a layered charge density wave (CDW) compound, EuTe$_\text{4}$. Using high-momentum-resolution X-ray diffraction performed on ultrathin flakes, we found two coexisting, incommensurate CDWs with slightly mismatched in-plane wavevectors. The interaction between these two CDWs leads to their joint commensuration with the high-symmetry lattice as well as a large moir\'{e} superstructure with an in-plane period of 13.6~nm. Due to different out-of-plane orders of the incommensurate CDWs, the moir\'{e} superstructure exhibits a clear thermal hysteresis, accounting for the large hysteresis observed in electrical resistivity and numerous metastable states induced by light or electrical pulses. Our findings pave the way for a new development in moir\'{e} engineering based on an incommensurate lattice. They further highlight the important role of interlayer ordering in determining the macroscopic properties of these stacked incommensurate structures.
\end{abstract}

\maketitle

\begin{figure*}[htb!]
\centering
\includegraphics[width=0.9\textwidth]{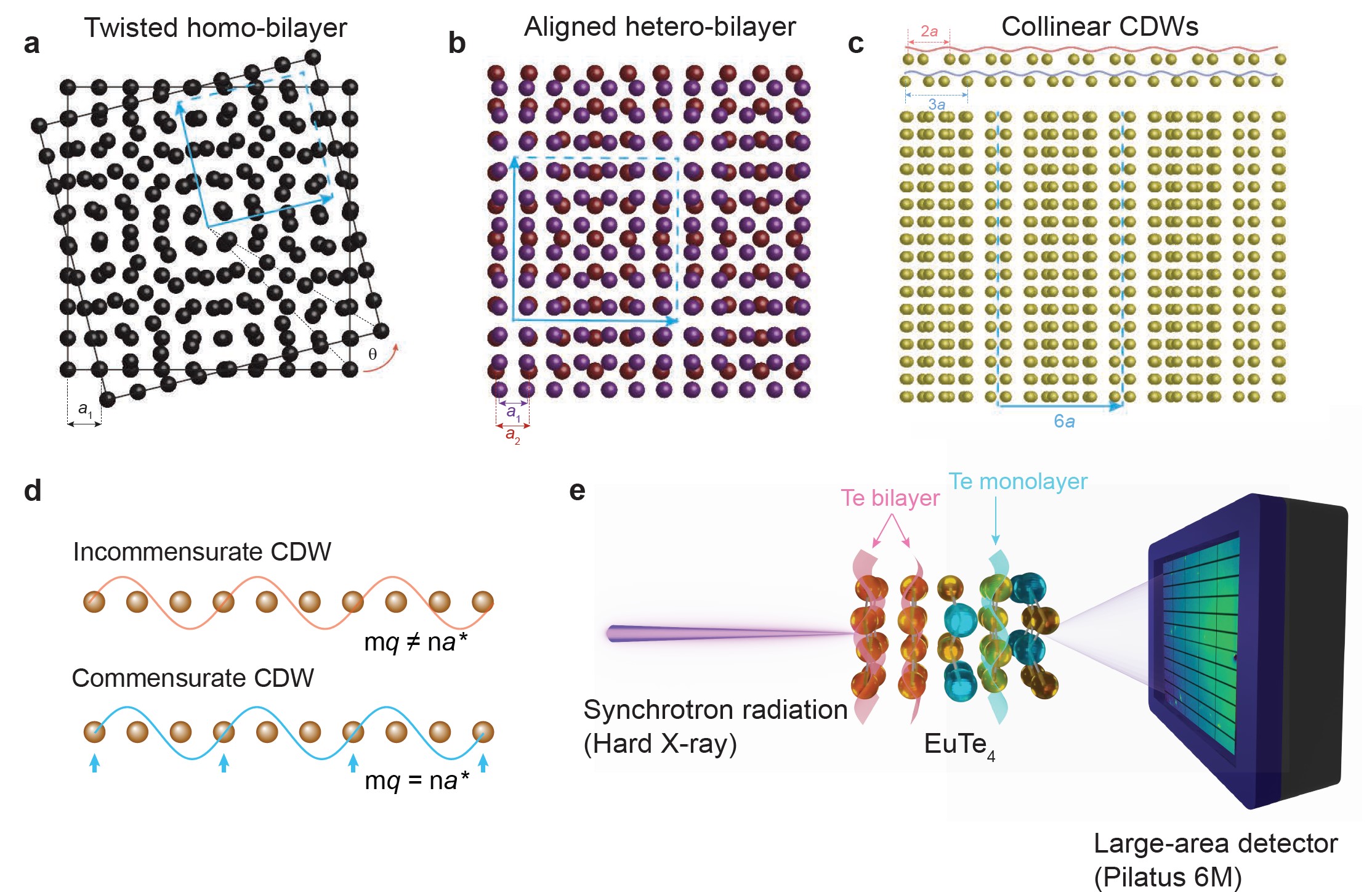}
\caption{\textbf{Moir\'{e} superlattice formed by stacking two CDWs.} \textbf{a}--\textbf{c}, Pathways towards the construction of a moir\'{e} superlattice: \textbf{a},~Stacking of twisted homo-bilayer, where two identical layers of crystals are stacked with a twist angle. \textbf{b},~Stacking of aligned hetero-bilayers, where two layers of different crystals with similar lattice parameters are stacked to form a moir\'{e} pattern. \textbf{c},~A new way of constructing moir\'{e} superlattice, where multiple charge density waves (CDWs) with slightly different wavevectors are aligned in parallel, forming a beating pattern and hence moir\'{e} superlattice. \textbf{d},~Comparison between incommensurate CDW (I-CDW) and commensurate CDW. For an I-CDW, the ratio between the wavevector and the lattice constant is an irrational number; for a commensurate CDW, it is a rational number. $m$ and $n$ are co-prime positive integers. \textbf{e},~Crystal structure of EuTe$_4$ and the schematic of the X-ray reciprocal space mapping experiment. The structure of EuTe$_4$ features alternating monolayer and bilayer Te square-net sheets. The synchrotron X-ray diffraction measurement was performed in transmission geometry. The crystal was rotated 360$^\circ$ continuously to cover the entire reciprocal space. The diffraction signals were detected by a large-area X-ray detector.}
\label{fig:Fig1}
\end{figure*}

Quasi-2D materials provide a rich platform for exploring a broad range of novel phenomena, exemplified by recent advances in moir\'{e} superstructures \cite{Cao2018,Mak2022}. Moir\'{e} systems are typically realized through the vertical stacking of 2D atomic layers with either a small twist angle $\theta$ and/or a small lattice mismatch ($a_1$, $a_2$, referring to the lattice constant of the two tetragonal/triangular/hexagonal stacking layers) \cite{Cao2018,Mak2022,Park2023,Jin2019,Zhang2017}, as depicted in Fig.~\ref{fig:Fig1}\textbf{a} and \textbf{b}. The resulting moir\'{e} period is determined by \cite{Mak2022}
\begin{equation}
    a_m \approx \left[\left(\frac{1}{a_1}\right)^2+\left(\frac{1}{a_2}\right)^2-\frac{2\cos{\theta}}{a_1a_2}\right]^{-1/2}.
\end{equation}
By tuning $a_m$, one can precisely tailor the macroscopic properties of these 2D vertical heterostructures. In particular, when $a_m$ significantly exceeds the atomic lattice constant, the Coulomb interactions ($U\sim a_m^{-\frac{1}{2}}$) between electrons within the same moir\'{e} period become comparable to or even surpass their kinetic energy ($E_k\sim a_m^{-2}$) (ref.~\cite{Mak2022}). This relative enhancement in electron-electron interactions can lead to a variety of insulating, superconducting, nematic, or magnetic states that are beyond the scope of single-particle physics in isolated 2D layers.

The vast majority of moir\'{e} systems are created out of crystalline or crystallographically commensurate 2D layers, such as graphene or transition metal dichalcogenides. On the other hand, materials with incommensurate orders intrinsically possess long-wavelength structural motifs, and they are the host to a number of important phenomena in solids, including translational symmetry breaking \cite{Axe1980}, superspace group formation \cite{deWolff:a19729}, density wave modulation and phase transition both in and out of thermal equilibrium \cite{Peierls1955,Gruner2018DensitySolids,Zong2021}, and excitation of exotic collective modes \cite{Pekker2015,wang_axial_2022,kim_observation_2023}. Given the interest in both the artificially stacked \textit{commensurate} structures to form a moir\'{e} pattern and the availability of naturally occurring \textit{incommensurate} orders, stacking two incommensurate lattices with different periods to form a moir\'{e} superstructure (Fig.~\ref{fig:Fig1}\textbf{c}) hence presents a promising avenue for exploring new types of moir\'{e} physics.

In this work, we report the observation of a large moir\'{e} superlattice realized by stacking two incommensurate charge density wave (I-CDW) orders in a quasi-2D material, EuTe$_4$. The two CDW orders, originating from the stacking of tellurium monolayers and bilayers within a unit cell, are aligned with the lattice through a joint commensuration condition $\mathbf{q}_1 + 2\mathbf{q}_{2,\text{in-plane}} = 2\textbf{b}^*$. Interestingly, the direct competition between the jointly commensurate I-CDWs and moir\'{e} relaxations results in a uniquely large hysteretic moir\'{e} superlattice, underlying the observed giant hysteresis in resistivity \cite{Lv2022UnconventionalWave}. Our findings present a new platform for realizing and manipulating the novel physics of hysteretic moir\'{e} superstructures.

CDW is one of the most prevalent and extensively studied orders in low-dimensional systems. The CDW wavevector $\mathbf{q}$ is oftentimes determined by the shape of the Fermi surface, so $\mathbf{q}$ can in general be incommensurate with respect to the underlying crystalline lattice (Fig.~\ref{fig:Fig1}\textbf{d}). In many incommensurate CDWs, $q \equiv|\mathbf{q}|$ is found to be temperature dependent, varying by a few percent from $T_\text{CDW}$ to $\sim 0$~K (refs.~\cite{Banerjee2013ChargeTellurides,DiCarlo1994,Feng2015,Shin2010,Ravy2006,Yue2020,Lee2020}). In some cases, if the ratio of $q$ to the lattice constant is close to an integer or a rational number consisting of small integers, the interaction between the two periodic structures can drive a transition to a commensurate CDW phase at a certain temperature below $T_\text{CDW}$, known as a ``lock-in'' transition \cite{Fleming1981,Fleming1985,Wilson1975}.

\begin{figure*}[htb!]
\centering
\includegraphics[width=\textwidth]{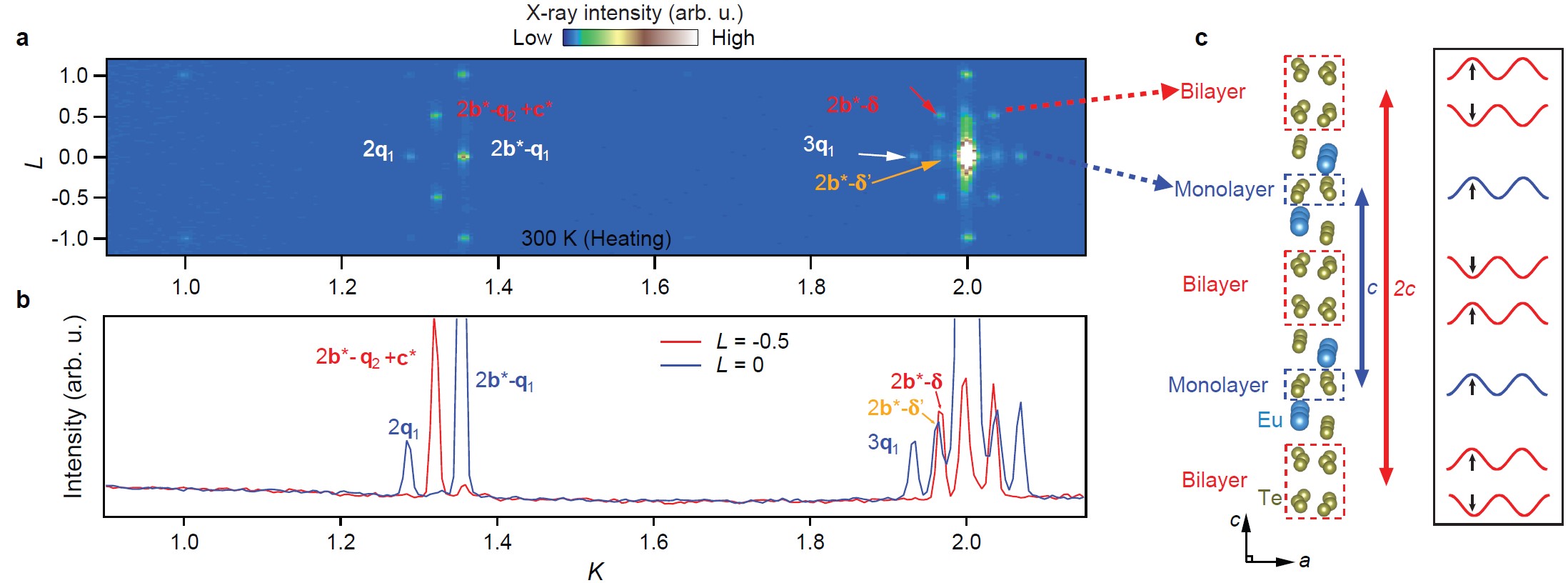}
\caption{\textbf{X-ray diffraction pattern and refined CDW structure of EuTe$_\text{4}$.} \textbf{a},~$(2,K,L)$ cut of x-ray diffraction (XRD) reciprocal space mapping at 300~K in the heating branch, featuring two sets CDW satellite peaks at integer and half-integer $L$ respectively. \textbf{b},~line-cut of the diffraction image in \textbf{a} along $L=0$ and $L=-0.5$. \textbf{c},~Schematic of charge density wave distortion configuration in EuTe$_4$ refined from a combination of powder XRD and reciprocal space mapping data. The integer $L$ and half-integer $L$ CDW satellites, respectively, correspond to CDW distortions in monolayer and bilayer Te sheets. The phase of bilayer CDW shifts by 180$^\circ$ in adjacent unit cells, forming a two-unit-cell superlattice and hence manifesting the half-integer CDW peaks in the out-of-plane direction.}
\label{fig:Fig2}
\end{figure*}

I-CDW modulation was recently observed in a quasi-2D layered Eu-based telluride EuTe$_4$, which immediately attracted considerable attention due to several notable anomalies  \cite{Wu2019LayeredSheets,Lv2022UnconventionalWave,Zhang2023Thermal4,Lv2024CoexistenceSemiconductor,Liu2024,zhangchen2022,Pathak2022,xiaokebin2024,ranjana2023}. First, the I-CDW order is associated with a giant thermal hysteresis, spanning from below 100~K to over 500~K, which represents the largest thermal hysteresis reported in crystalline solids \cite{Wu2019LayeredSheets,Lv2022UnconventionalWave,Zhang2023Thermal4}. Second, the measured in-plane modulation wave vector $\mathbf{q}_{\text{in-plane}}$ is constant within a temperature range of 50--400~K (ref.~\cite{Lv2022UnconventionalWave}), and the hysteresis does not result from the well-known ``lock-in'' transition, even though $\mathbf{q}_{\text{in-plane}} = 0.643(3)\mathbf{b}^*$ is close to $(2/3)\mathbf{b}^*$, where $\mathbf{b}^*$ is the reciprocal lattice vector. Third, the Fermi surface is fully gapped in the I-CDW state despite the absence of a perfect Fermi surface nesting condition \cite{Lv2022UnconventionalWave,Zhang2023Thermal4,zhangchen2022}.

Structure-wise, as the only Eu-based layered telluride \cite{Yumigeta2021AdvancesSynthesis}, EuTe$_4$ hosts a rare coexistence of tellurium monolayers and bilayers within a single unit cell (Fig.~\ref{fig:Fig1}\textbf{e}), suggesting the presence of two distinct types of CDWs. Indeed, previous time- and angle-resolved photoemission spectroscopy (tr-ARPES) measurements have observed the characteristic energy gaps of monolayer and bilayer CDWs \cite{Lv2024CoexistenceSemiconductor}. In particular, the observation of momentum-dependent gap renormalization of monolayer CDWs indicates additional interlayer interactions resulting from direct charge transfer between the nominally charge-neutral Te monolayer and bilayers, an interpretation that is also supported by density-functional theory (DFT) calculations. This additional monolayer-bilayer CDW interaction was understood as the main driving force of the observed semiconducting CDW state \cite{Lv2024CoexistenceSemiconductor}. On the other hand, because of the direct charge transfer, one would expect the Te monolayer and bilayer to exhibit distinct Fermi surfaces, and thereby different CDW modulations. X-ray diffraction (XRD) is an ideal technique for probing the modulation wave vector \cite{Warren1990X-rayDiffraction}. However, for quasi-2D layered materials, crystalline defects such as stacking faults usually obscure the identification of intrinsic spatial modulations along the $c$-axis, and this explains why previous XRD measurements of EuTe$_4$ bulk materials revealed multiple diffraction peaks and lines along the $L$ direction \cite{Lv2022UnconventionalWave,Rathore2023EvolutionEuTe4} (see Supplementary Information for further details).

\begin{figure*}[htb!]
\centering
\includegraphics[width=0.95\textwidth]{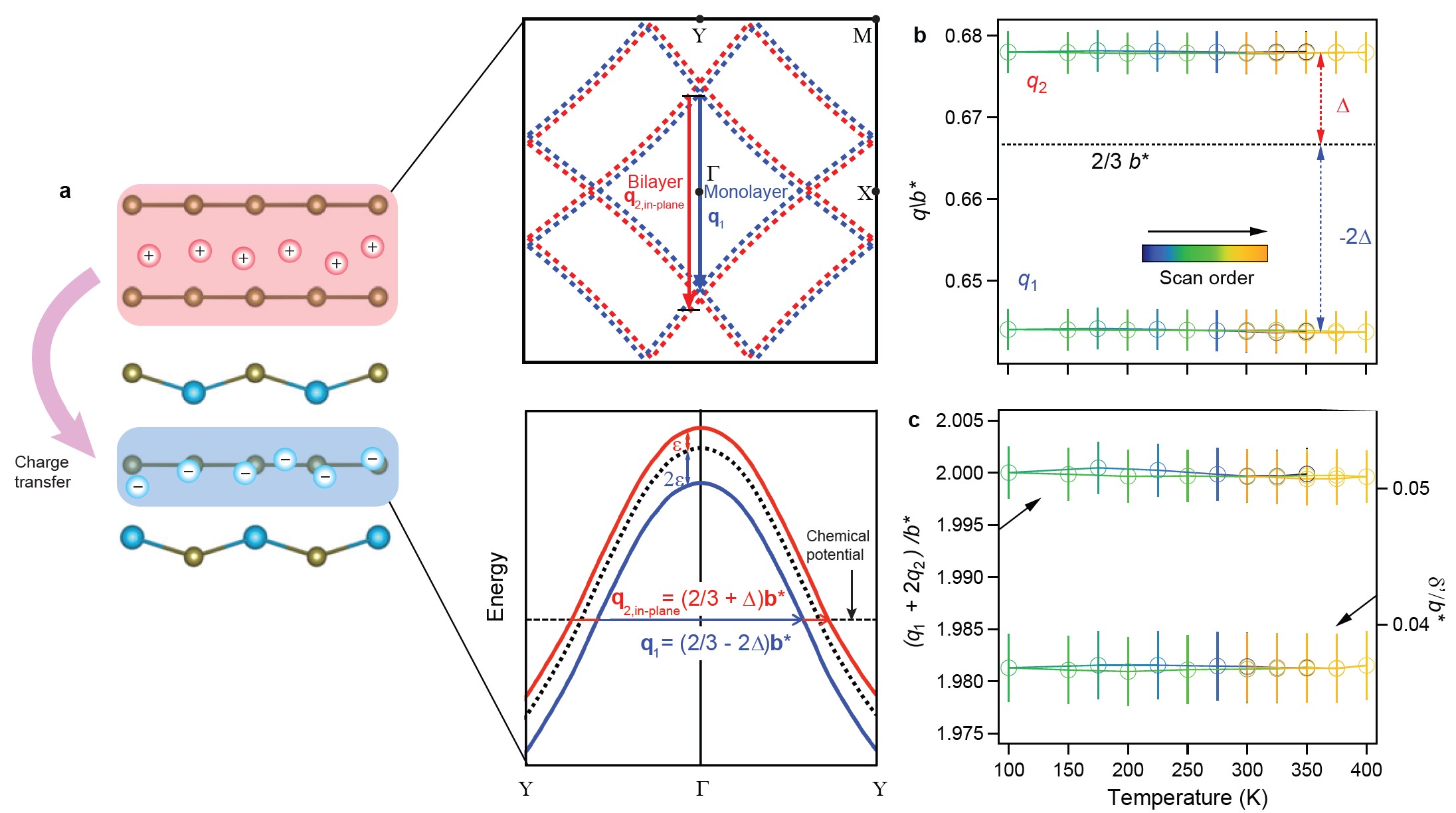}
\caption{\textbf{Jointly commensurate CDW orders induced by charge transfer.} \textbf{a},~Schematics of charge transfer in real space (left) and consequent band structure shifts in reciprocal space, indicated in the schematic Fermi surface (top right) and schematic band dispersion along the Y-$\Gamma$-Y direction (bottom right) of the non-CDW state of EuTe$_4$ from tight-binding calculation. The charge transfer process can be understood as doping the nominally charge-neutral square-like sheet in the Te bilayer with 1~hole while doping monolayer sheets with 2~electrons. This results in an energy shift in monolayer (solid blue curve) and bilayer (solid red curve) bands from the nominally charge-neutral band position (dotted black curve) in opposite directions. Taking commensurate wavevector $\mathbf{q}=2/3\mathbf{b}^*$ as a starting point, the in-plane components of wavevectors in monolayer and bilayer will respectively change by $-2\Delta$ and $\Delta$ (in units of $\mathbf{b}^*$), respectively, due to the change in the geometry of electronic bands. \textbf{b},~CDW in-plane wavevector amplitudes of monolayer ($q_1$) and bilayer ($q_2$) as a function of temperature, showing temperature invariance across a thermal loop spanning from 100~K to 400~K. \textbf{c},~Temperature dependence of the joint commensuration condition $q_1+2q_2$ and the reconstructed moir\'{e} superlattice wavevector amplitude ($\delta'$), both showing negligible temperature-dependent changes within experimental uncertainties.}
\label{fig:Fig3}
\end{figure*}

To minimize the impact of stacking faults, we fabricated high-quality ultrathin flakes with thicknesses ranging from approximately 20 to 50~nm (see Fig.~S2). Leveraging the superior flux and resolution of synchrotron-based X-ray diffraction, we successfully discerned the out-of-plane density wave modulation periods. Figure~\ref{fig:Fig2}\textbf{a} shows a representative XRD pattern in the $(2,K,L)$ plane at 300~K in the heating branch. Remarkably, compared to bulk data shown in Fig.~S1, the measured diffraction pattern is greatly improved. Particularly notable are the well-defined and sharp diffraction peaks observed at $c^*$ and $0.5c^*$, facilitating the identification of intrinsic out-of-plane CDW modulations. Specifically, one can use a minimum of two modulation vectors to describe most of the satellite diffraction peaks:
\begin{align}
    \mathbf{q}_1 &= 0.644(5)\mathbf{b}^*,\\
    \mathbf{q}_2 &= 0.678(5)\mathbf{b}^*+0.5\mathbf{c}^*,
\end{align}
as marked in Fig.~\ref{fig:Fig2}\textbf{a} and \textbf{b}. The unambiguous identification of both $\mathbf{q}_1$ and $\mathbf{q}_2$ provides direct evidence for coexisting CDW orders in EuTe$_4$. One can see that both modulations are not commensurate with the lattice under a small integer denominator, confirming their incommensurate nature. Furthermore, it is worth noting that the lattice distortion does not conform strictly to a sinusoidal form, resulting in pronounced diffraction intensities in the second and third harmonics of the superlattice peaks \cite{Overhauser1971ObservabilityDiffraction}.

The identification of dual modulations in ultrathin flakes provides hitherto the strongest evidence for the existence of monolayer and bilayer CDWs that have been suspected in EuTe$_4$ (ref.~\cite{Lv2024CoexistenceSemiconductor}). Previous bulk crystal XRD refinement \cite{Wu2019LayeredSheets} revealed a period of $c$ and $2c$ for monolayer and bilayer Te distortion, respectively, as shown in Fig.~\ref{fig:Fig2}\textbf{c}. In light of our present ultrathin single-crystal XRD measurements, we conclude that $\mathbf{q}_1$ and $\mathbf{q}_2$ correspond to monolayer and bilayer CDWs, respectively. Notably, the in-plane modulation wavevectors of $\mathbf{q}_1$ and $\mathbf{q}_2$ only differ by a small value, $\delta_\text{in-plane} = 0.034b^*$. The vertical stacking of these two incommensurate CDW orders presents an ideal platform for exploring long-sought-after incommensurate order moir\'{e} physics within a single crystal itself. Indeed, the XRD data reveals clear and sharp diffraction peaks at $2\mathbf{b}^*-\pmb{\updelta}$, directly evidencing the formation of a moir\'{e} pattern along the $b$ direction with a long spatial period of $2\pi/\delta_\text{in-plane}\approx 13.6$~nm. 
As a reference, one needs a small twist angle of $1.26^{\circ}$ or a tiny mismatch of 6.6~pm in lattice constant to achieve the same moir\'{e} period by stacking two layers, where we assumed a lattice parameter of $\sim 3$~{\AA}, which is comparable to the commonly used atomic layers, like graphene and transition metal dichalcogenides.

Observing a room-temperature I-CDW moir\'{e} superstructure with such a long spatial period is rare, 
and it exists in EuTe$_4$ for several reasons. First of all, as mentioned earlier, the I-CDW states in EuTe$_4$ stem from the nominally charge neutral Te monolayer and bilayer square nets. This unique structure suggests that the two stacked I-CDW states likely exhibit in-plane-collinear modulations with closely matched nesting vectors. Moreover, additional charge transfer from the bilayer to the monolayer (illustrated in Fig.~\ref{fig:Fig3}\textbf{a}) results in a slight variation in the exact wave vector, crucial for defining the moir\'{e} lattice. More precisely, one expects a relatively smaller wave vector for the monolayer CDW due to electron doping, consistent with the X-ray diffraction (XRD) findings in Fig.~\ref{fig:Fig1}. The above analysis not only underscores the pivotal role of interlayer charge transfer in forming the I-CDW moir\'{e} pattern but also proposes an effective method to tune the moir\'{e} period through electronic gating that is expected to precisely control the wavevector in atomically-thin systems.

Having established the existence of I-CDW moir\'{e} superstructures, we now investigate their effects on the unconventional giant thermal hysteresis in EuTe$_4$ (ref.~\cite{Lv2022UnconventionalWave}). We first studied the temperature dependence of the superstructures. Figure~\ref{fig:Fig3}\textbf{b} depicts the temperature dependence of $\mathbf{q}_1$ and $\mathbf{q}_{2,\text{in-plane}}$ from 100~K to 400~K. Surprisingly, both $\mathbf{q}_1$ and $\mathbf{q}_{2,\text{in-plane}}$ exhibit no discernible change within our experimental resolution, which contrasts with the typical temperature-dependent spatial modulation observed in many I-CDW materials \cite{Ru2008,Banerjee2013ChargeTellurides}. The robustness of these incommensurate wave vectors over such a wide temperature range suggests the existence of an intrinsic joint ``lock-in'' mechanism. To elucidate this mechanism, we revisit the wave vectors. We begin by assuming that each Te monolayer and bilayer hosts an imaginary commensurate CDW modulation of $(2/3)b^*$, as suggested by DFT calculations \cite{Wu2019LayeredSheets}. When these layers stack together, additional charge transfer alters the electron filling. Specifically, if each Te layer in the bilayer donates one electron, the Te monolayer receives two electrons, leading to a downshift and upshift of the inner chemical potential of $\epsilon$ and $2\epsilon$, respectively. Very importantly, because the Te monolayer and bilayer have similar electronic structure with nearly linear dispersed $p_x/p_y$ bands near the chemical potential, this charge transfer eventually induces changes in the nesting vectors: $\mathbf{q}_1 = (2/3 - 2\Delta)\mathbf{b}^*$ for the monolayer CDW and $\mathbf{q}_{2,\text{in-plane}} = (2/3 + \Delta)\textbf{b}^*$ for the bilayer CDW, as illustrated in Fig.~\ref{fig:Fig2}\textbf{a}. In this situation, despite being incommensurate, $\mathbf{q}_1$ and $\mathbf{q}_{2,\text{in-plane}}$ are locked together with the lattice via the relationship $\mathbf{q}_1 + 2\mathbf{q}_{2,\text{in-plane}} = 2\textbf{b}^*$. This joint ``lock-in'' relation is directly verified by our temperature-dependent measurements shown in Fig.~\ref{fig:Fig3}\textbf{c}.

It is the joint ``lock-in'' mechanism that accounts for the previously reported puzzling observation that the incommensurate CDW wave vector in this material is temperature independent\cite{Lv2022UnconventionalWave}. On the other hand, while the two stacked I-CDWs are jointly locked, the resulting moir\'{e} pattern still has an incommensurate in-plane period, i.e., $\delta_\text{in-plane} = 0.034b^*$.  In principle, interlayer interactions have the potential to reshape the incommensurate moir\'{e} structure, leading to the formation of intriguing commensurate moir\'{e} domains with different crystalline symmetry and electronic structures \cite{Woods2014,Yoo2019,Wijk_2015}. Referring back to the XRD data in Fig.~\ref{fig:Fig2}\textbf{a}, in addition to the anticipated strong I-CDW and resulting moir\'{e} diffraction peaks at $2\mathbf{b}^*\pm\pmb{\updelta}$, two extra satellite peaks are observed at $L = 0$, labeled as $2\mathbf{b}^*\pm\pmb{\updelta'}$, indicated by the yellow arrow. Upon closer examination of the corresponding high-resolution X-ray intensity profile in Fig.~\ref{fig:Fig2}\textbf{b}, it becomes evident that ${\delta'} = 0.037b^*$ is slightly larger than $\delta_\text{in-plane}$. Specifically, $\delta'$ corresponds to a real-space period of $2\pi/\delta' = 27$~unit cells along the $b$-axis, confirming the commensurate moir\'{e} reconstructions induced by interlayer coupling.

\begin{figure*}[htb!]
\centering
\includegraphics[width=1\textwidth]{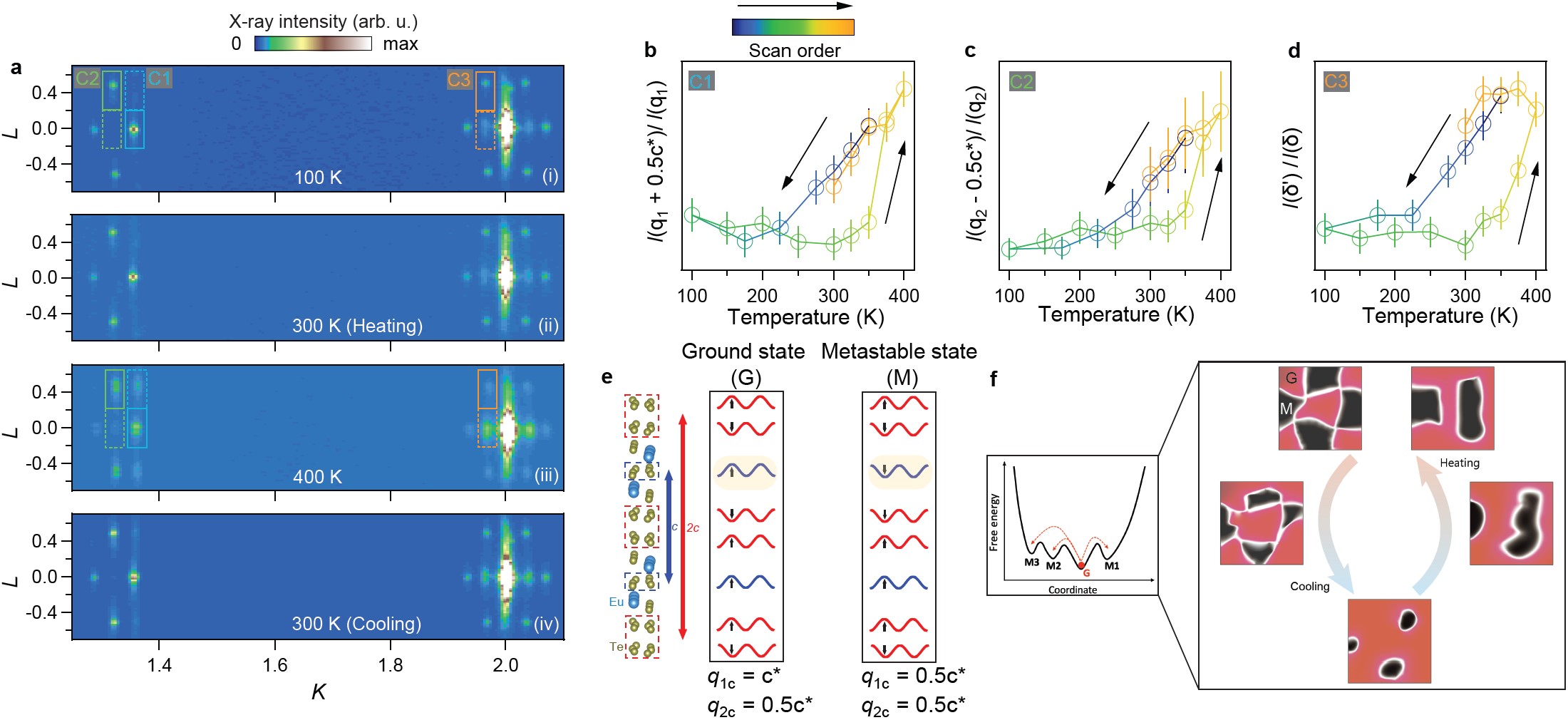}
\caption{\textbf{Giant thermal hysteresis in EuTe$_\text{4}$ due to metastable domains of CDW stacking disorder.} \textbf{a},~$(2, K, L)$ cuts of reciprocal space mapping at different temperatures in a thermal cycle: 350~K \textrightarrow 100~K \textrightarrow 400~K \textrightarrow 300~K [see (i)--(iv)]. \textbf{b}--\textbf{d},~Ratio of intensity in the region of interest (ROI) (dashed over solid for each pair) indicated in \textbf{a} for different $K$ values, labeled as C2, C1, and C3, respectively. The dashed and solid ROIs represent regions where satellite peaks are forbidden and allowed respectively in the ground state without disorder. These ratios reflect the degree of stacking disorder due to thermal excitations. \textbf{e},~Schematics of CDW configurations for ground state (G) and metastable states (M). The up and down arrows indicate the phase of CDWs with respect to the lattice. The yellow shade highlights the difference in the monolayer CDW phase between the two configurations. Only one type of metastable state is demonstrated here for simplicity. A more detailed discussion of metastable state configurations can be found in Supplementary Information. \textbf{f}, Real-space schematics of ground state (G, red) and metastable states (M, black) domains, which evolve over the temperature cycle; the schematics are inferred from the diffraction data. The white boundaries indicate domain walls. The specific configuration of domain distribution may vary throughout the thickness of the sample while the ratio between the area of G and M domains remains similar through out the sample.}
\label{fig:Fig4}
\end{figure*}

We would like to point out the observation of $\delta'$ peaks at $L = 0$ planes are unexpected if we only consider a stacking of monolayer order with period $c$ and bilayer order with period $2c$. In this case, the resultant beating pattern would have a $2c$ period and thus should only appear at half-integer $L$. To resolve this contraction, the interaction, and hence lattice reconstruction, between monolayer and bilayer Te sheets should be taken into consideration. The $\delta'$ peaks in the $L=0$ plane are no longer forbidden if the reconstructed superlattice does not strictly follow a $2c$ period. 
Conversely, the presence of $\delta'$ peaks demonstrates that the moir\'{e} reconstruction primarily disrupts the long-range out-of-plane phase coupling of monolayer and/or bilayer CDWs, meaning that the relative phases between the CDW orders in adjacent unit cells no longer strictly follow the out-of-plane periodicity in the ground state. Note that previous studies have linked the out-of-plane phase changes to the mechanism underlying giant thermal hysteresis and light-induced metastable states, implying a close relation with the observed commensurate moir\'{e} reconstruction \cite{Lv2022UnconventionalWave,Lv2024CoexistenceSemiconductor,Liu2024}. 

To understand their correlation, we conducted XRD measurements across a complete thermal cycle from 100~K to 400~K. Figure~\ref{fig:Fig4}\textbf{a} presents four representative XRD maps at 100~K, 300~K (cooling branch), 300~K (heating branch), and 400~K. We mainly focused on the satellite peaks, which are classified into two categories: the ground-state satellite peaks, including the satellite peaks of monolayer and bilayer CDWs and the moir\'{e} diffraction peaks $\delta$ (marked by solid rectangles, with different colors at different $K$ values, in Fig.~\ref{fig:Fig4}\textbf{a}(i)), and the forbidden satellite peaks, which refer to the originally forbidden peaks lightened up by moir\'{e} reconstruction or out-of-plane phase changes, including $\pmb{\updelta'}$, $\mathbf{q}_1 \pm 0.5\mathbf{c}^*$, $\mathbf{q}_2 \pm 0.5\mathbf{c}^*$, as highlighted by dashed rectangles in Fig.~\ref{fig:Fig4}\textbf{a}(i). Upon comparison of the XRD maps at 100~K and 400~K, it is evident that the ground-state satellite peaks dominate over the forbidden peaks at 100~K, whereas the opposite trend is observed at 400~K. This inverse temperature dependence points to direct competition between the jointly-locked intralayer I-CDWs and interlayer moir\'{e} reconstructions. Remarkably, in accordance with the thermal hysteresis phenomenon, the forbidden peaks also show a history-dependent behavior, as summarized in Fig.~\ref{fig:Fig4}\textbf{b}--\textbf{d}. Specifically, the intensity of forbidden peaks at 300~K is stronger in the cooling branch compared to the heating branch. In other words, the strength of I-CDWs is more pronounced in the heating branch compared to the cooling branch, giving rise to the observed hysteretic resistivity and energy gaps.

The above observations in Fig.~\ref{fig:Fig4}a--d suggest a microscopic origin of the reported thermal hysteresis in electrical resistivity as well as pulse-induced metastable states in EuTe$_4$ \cite{Lv2022UnconventionalWave,Liu2024,Venturini2024}. Recall that the out-of-plane period of the monolayer (or bilayer) CDW is $c$ (or $2c$) in the ground state; in other words, the phase between two adjacent Te monolayer (or bilayer) is $\theta_1 =0$ (or $\theta_2=\pi$) \cite{Wu2019LayeredSheets}. Given the van der Waals nature of the layered compound where in-plane couplings are much stronger than out-of-plane couplings, metastable states can be formed if $(\theta_1, \theta_2)$ takes on values other than $(0,\pi)$. If we restrict $\theta_1,\theta_2$ to either 0 or $\pi$, there are three categories of metastable states, as enumerated in Figs.~\ref{fig:Fig4}\textbf{e} and S3. Under this framework, Fig.~\ref{fig:Fig4}\textbf{f} illustrates the real-space schematic of the ground state and metastable state during the temperature cycle.  

The nature of the joint lock-in relation and metastable states with different out-of-plane orders is further substantiated by our symmetry analysis. In particular, we focus on the interlayer couplings, including the ones within one unit cell and between different unit cells. Within one unit cell, we have three Te layers with CDWs, whose in-plane momenta are denoted by $\mathbf{q}_{1,2,3}$. 
CDWs within individual layers spontaneously break the local translation symmetry of corresponding layers.
For the global translation symmetry of the unit cell, two scenarios can be present. 
If the global translation symmetry is also spontaneously broken, then there will be a net incommensurate CDW order. 
If the global translation symmetry is preserved, then the total momentum is conserved up to a 
reciprocal vector of individual layers $\mathbf{q}_1+\mathbf{q}_2+\mathbf{q}_3 =\mathbf G$. 
The interlayer coupling within the Te bilayer is so strong that the two layers share the same CDW $\mathbf{q}_{2}=\mathbf{q}_{3}$. As a result, we arrive at the joint lock-in relation $\mathbf{q}_1+2\mathbf{q}_2=\mathbf G$. As elaborated in the Supplementary Information, through mean-field analysis, we find that strong CDW-lattice interaction is required to preserve the global translational symmetry. In addition, the joint lock-in relation is fulfilled in a wide temperature in EuTe$_4$ also thanks to the fact that the nesting vector of each Te layer is very close to $2/3b^*$. 

By employing advanced high-flux and high-resolution XRD measurements on exfoliated ultrathin flakes of EuTe$_4$, we provide direct evidence for the coexistence of monolayer and bilayer CDWs, which are characterized by incommensurate modulation vectors, $\mathbf{q}_1 = 0.644(5)\mathbf{b}^*$ and $\mathbf{q}_2 = 0.678(5)\mathbf{b}^* +0.5\mathbf{c}^*$, respectively. The slight difference in the in-plane component of $q_1$ and $q_2$ naturally gives rise to a moir\'{e} superstructure with an incommensurate modulation vector, $\delta_\text{in-plane} = 0.034b^*$, corresponding to a spatial period as large as approximately 13.6~nm. Notably, both $q_1$, $q_2$, and $\delta$ are robust against temperature, evidencing a joint ``lock-in'' of monolayer and bilayer CDWs under relation $\mathbf{q}_1 + 2\mathbf{q}_{2,\text{in-plane}} = 2\textbf{b}^*$. On the other hand, interlayer coupling and thermal fluctuations further reconstruct the incommensurate moir\'{e} superstructure, leading to the formation of hysteretic commensurate moir\'{e} domains with varying out-of-plane phase orderings of monolayer and bilayer CDWs. The observed inverse temperature dependence of diffraction intensities of $\delta$ and $\delta'$ illustrates a direct competition between the jointly locked intralayer I-CDWs and interlayer moir\'{e} reconstructions. This competition is pivotal in understanding and manipulating the anomalously large hysteresis observed in EuTe$_4$. Our findings provide critical insights into the unique properties of EuTe$_4$, such as its resistivity hysteresis and the presence of multiple metastable states. This discovery of a large and stable moir\'{e} superstructure opens new avenues for exploring moir\'{e} physics in layered materials via stacked incommensurate orders.

\section{Methods}
\subsection{Material preparation}
EuTe$_4$ single crystals were synthesized via a solid-state reaction with Te as the flux. Stoichiometric Eu lumps (99.999\%) and Te granules (99.999\%) were mixed with a ratio of approximately 1:15. The total weighted starting materials were sealed in an evacuated fused silica tube under high vacuum ($10^{-5}$ mbar) followed by heating at $850^\circ$C for two days in a muffle furnace. The furnace was slowly cooled to $415^\circ$C over 100 hr, held at this temperature for one week, and then decanted using a centrifuge \cite{Wu2019LayeredSheets}. Unlike $R$Te$_3$, EuTe$_4$ single crystals are stable under ambient conditions. They are planar-shaped with dark and mirror-like surfaces, where the surface area is up to 1~mm$^2$ and the thickness up to 0.2~mm.

EuTe$_4$ flakes ($\sim$ 20 to 50 nm thick) used in the experiments were mechanically exfoliated from the bulk crystals and then transferred to amorphous glass cover slides. The flakes are oriented in the $(0 0 1)$ plane with a typical surface area of about 300$\times$300~$\upmu$m$^2$.

\subsection{X-ray reciprocal space mapping}

X-ray reciprocal space mapping experiments were carried out at the QM2 beamline at the Cornell High Energy Synchrotron Source (CHESS) \cite{chess}. The incident X-ray energy of 9.5~keV was selected using a double-bounce diamond monochromator. A stream of cold-flowing nitrogen or helium gas was used to cool the sample. The diffraction experiment was conducted in transmission geometry using a 6-megapixel photon-counting pixel-array detector with a silicon sensor layer (Pilatus 6M). Data were collected in 360$^\circ$ sample rotations with a step size of 0.1$^\circ$. After data acquisition, all images are stacked together to form 3D data labeled with three diffraction angles. A least-squares fitting is performed on all identified peaks to generate an orientation matrix that maps the data from the diffraction angle space to the $(H,K,L)$ space.

\section{Additional Information}

\noindent \textbf{Acknowledgments.}~The authors thank Riccardo Comin, Anshul Kogar, Honglie Ning, Kyoung Hun Oh, and Darius Shi for helpful discussions. The work at MIT was supported by the U.S. Department of Energy, the BES DMSE (data collection and analysis) and the Gordon and Betty Moore Foundation’s EPiQS Initiative grant GBMF9459 (manuscript writing). Research conducted at the Center for High-Energy X-ray Science (CHEXS) is supported by the National Science Foundation (BIO, ENG and MPS Directorates) under award DMR-2342336. B.L. acknowledges support from the Ministry of Science and Technology of China (Grant No.~2023YFA1407400), the National Natural Science Foundation of China (Grant No. 12374063), and the Shanghai Natural Science Fund for Original Exploration Program (Grant No.~23ZR1479900). N.L.W. acknowledges support from the National Natural Science Foundation of China (Grant No. 12488201), and the National Key Research and Development Program of China (Grant Nos.~2024YFA1408700, 2022YFA1403901). D.W. acknowledges support from the National Key Research and Development Program of China (Grant No.~2024YFA1408700). N.F.Q.Y. acknowledges support from the National Natural Science Foundation of China (Grant No.~12174021).\\

\noindent \textbf{Author contributions.}~B.Q.L., Y.S., and A.Z. conceived the study. B.Q.L., Y.S., A.Z., S.S., and J.P.C.R. performed the X-ray diffraction measurements. Q.L., D.W., and N.L.W. synthesized, characterized, and prepared the EuTe$_4$ crystals. S.S. and J.P.C.R. maintained and set-up the synchrotron end-station at CHESS. N.F.Q.Y developed the mean-field theory. B.Q.L., Y.S., and A.Z. analyzed the data with the help of J.L., Z.N., and S.M. B.Q.L., Y.S., A.Z., and N.F.Q.Y. wrote the manuscript with critical input from N.G. and all other authors. The work was supervised by N.G.

\end{document}